\documentclass[11pt,twoside]{article}

\usepackage{asp2004}
\usepackage{epsf}
\usepackage{psfig}
\usepackage{lscape}

\begin{document}
\title{The neutral outflowing disks of some Magellanic Cloud B[e] supergiants} 
\author{Michaela Kraus}   
\affil{Sterrekundig Instituut, Utrecht University, Princetonplein 5, 3584 CC Utrecht, The Netherlands}   
\author{Marcelo Borges Fernandes}   
\affil{Observat\'{o}rio do Valongo (UFRJ), Ladeira do Pedro Ant\^onio 43, 20080-090 Rio de Janeiro, Brazil}
\author{Diana Andrade Pilling \& Francisco X. de Ara\'ujo}   
\affil{Observat\'{o}rio Nacional-MCT, Rua General Jos\'{e} Cristino 77, 20921-400 S\~{a}o Cristov\~{a}o, Rio de Janeiro, Brazil}

\begin{abstract} 
      We report on the detection of [O{\sc i}] emission lines in the
      high-resolution optical spectra of several Magellanic Cloud (MC) 
      B[e] supergiants, which we took with FEROS at the ESO 1.52m 
      telescope in La Silla (Chile). In addition, we model the [O{\sc i}]
      line luminosities and show that the best location for the neutral 
      oxygen material is the outflowing disk. In order to reproduce the 
      observed line luminosities, we conclude that the disks must be 
      neutral already very close to the stellar surface.
\end{abstract}

%%% MAIN BODY OF TEXT GOES HERE. CONSULT "INSTRUCTIONS FOR AUTHORS USING
%%% LATEX2E MARKUP", SECTIONS 2.3-2.6 FOR HELP WITH EQUATIONS, FIGURES,
%%% AND TABLES.

\section{Introduction}   %%% Top level section head (remove "%" symbol)

      The existence of disks around B[e] supergiants seems nowadays to be well
      established. To date it is, however, not clear how close to the
      surface of a hot star the
      material can drop neutral to allow molecule and dust formation.
      The [O{\sc i}] emission lines in the optical spectra of these stars
      (see Fig.\,1)
      are a perfect tracer of neutral material due to the equal ionization
      potential of H and O. They therefore mirror the neutral material in
      the circumstellar disk.

\section{The outflowing disk model}

      For our model calculations we assume that the [O{\sc i}] emitting
      regions are completely neutral in hydrogen. The free electrons
      necessary to collisionally excite O{\sc i} are provided by
      elements with much lower ionization potential, i.e Fe, Ca, Mg, etc.
      We use oxygen abundances of $1/3$ and $1/4$ solar for the LMC and
      SMC, respectively. The radial density distribution
      follows from the equatorial mass loss rate, 
      $\dot{M}_{\rm d}$, and terminal velocity, $v_{\infty}$, 
      via the equation of mass continuity; $N_{\rm e}$ is given in terms
      of $N_{\rm H}$.
                                                                                
\begin{figure}[!th]
\plotone{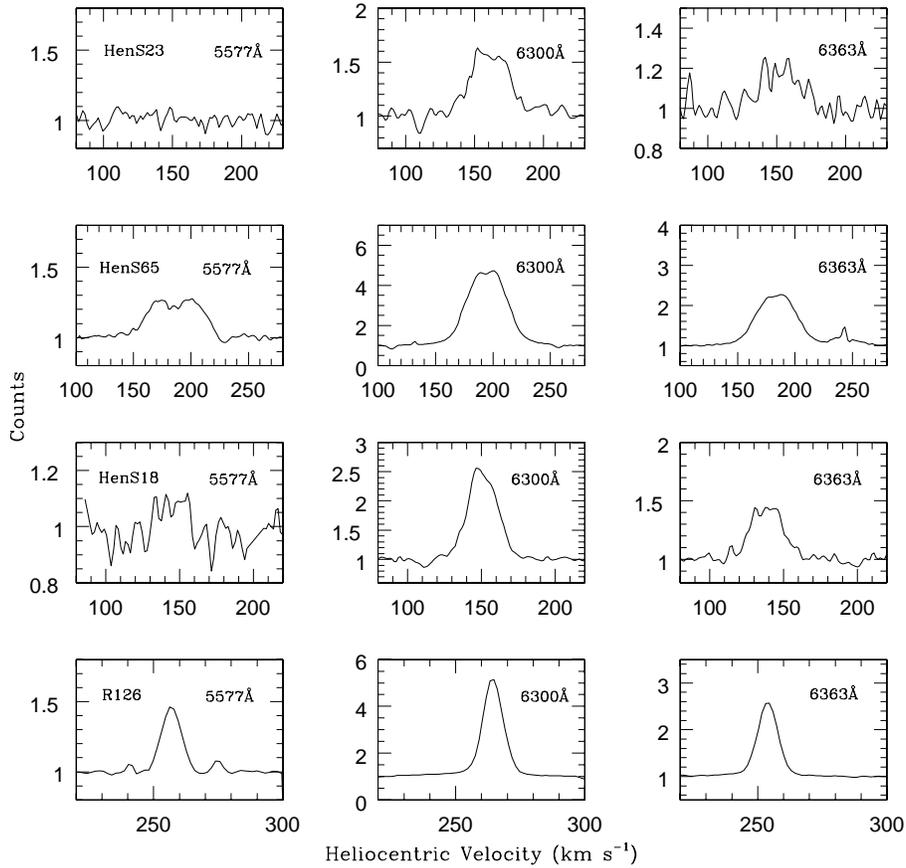}
\caption{The line profiles of the [O{\sc i}] lines
          arising in the optical spectrum for several Magellanic
          Cloud B[e] supergiants. The data were taken with FEROS.}
\end{figure}

      To calculate the level population we solve the statistical equilibrium
      equations in a 5-level atom; the forbidden lines are optically thin,
      simplifying the line luminosity calculation.
      The best-fit model parameters are summarized in Table\,1.
      Fixed parameters during calculations are: $v_{\infty} = 30$\,km/s
      (as suggested from the line wings of the [O{\sc i}] lines, Fig.\,1), 
      $T_{\rm e}\simeq 8000$\,K
      was found to be the best temperature to minimize $\dot{M}$, and the disk
      opening angle is set to 23$^{\circ}$.

%\begin{figure}[!th]
%\plotone{FIGOI.ps}
%\caption{The line profiles of the [O{\sc i}] lines
%          arising in the optical spectrum for several Magellanic
%          Cloud B[e] supergiants. The data were taken with FEROS.}
%\end{figure}

\begin{table}[!ht]
  \caption{Model parameters for the [O{\sc i}] line luminosity
         calculations. In the case of R\,126 we deal
         with two different scenarios (see Fig.\,3). The electron densities
         result from fitting the observed [O{\sc i}] line ratios.}
    \smallskip
      \begin{center}
        {\small
         \begin{tabular}{lcccc}
         \tableline
         \noalign{\smallskip}
       Star & $\dot{M}_{\rm d}$ & $\dot{M}_{\rm w}$ & $N_{\rm e}/N_{\rm H}$  \\
       & [$M_{\odot}$yr$^{-1}$sterad$^{-1}$] & [$M_{\odot}$yr$^{-1}$sterad$^{-1}$] &    \\
         \noalign{\smallskip}
         \tableline
         \noalign{\smallskip}
         Hen S 18 & $7.0\times 10^{-3}$ & & $6.0\times 10^{-8}$  \\
         Hen S 23 & $1.0\times 10^{-4}$ & & $4.3\times 10^{-5}$  \\
         Hen S 65 & $3.0\times 10^{-2}$ & & $8.9\times 10^{-9}$  \\
         \noalign{\smallskip}
         \tableline
         \noalign{\smallskip}
         R\,126 [OI] & $2.5\times 10^{-3}$ & & $2.0\times 10^{-6}$ \\
         R\,126 [OII] & $2.5\times 10^{-3}$ & & $1.0$ \\
         \noalign{\smallskip}
         \tableline
         \noalign{\smallskip}
         R\,126 [OI] & $3.6\times 10^{-3}$ & & $4.0\times 10^{-7}$ \\
         R\,126 [OII] & & $2.5\times 10^{-6}$ & 1.0 \\
         \noalign{\smallskip}
         \tableline
         \end{tabular}
        }
     \end{center}
  \end{table}

\begin{figure}[!ht]
\plotone{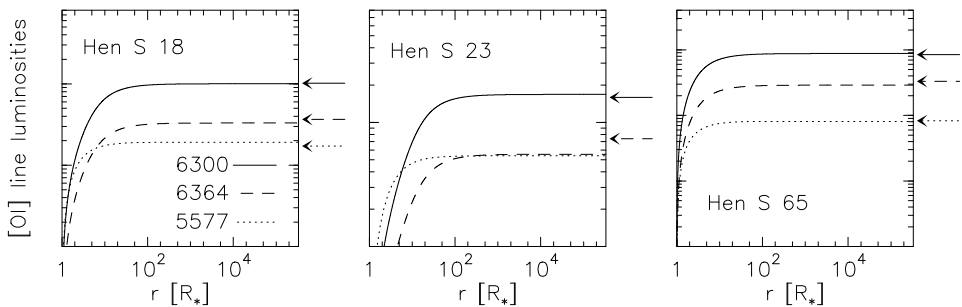}
\caption{Calculated increase in [O{\sc i}] line
          luminosities arising in the disks of the 
          different SMC B[e] supergiants.
          The arrows indicate the observed values.}
\end{figure}

\section{The SMC B[e] supergiants}

      For the SMC B[e] supergiants the [O{\sc i}] lines were (besides Fe)
      the only detectable forbidden emission lines, indicating
      that the emission region of the forbidden lines must be of rather high
      density and low ($\leq 10\,000$\,K) temperature.
      Fig.\,2 shows the increase in integrated line luminosity as a function
      of distance from the star. For comparison, the observed line
      luminosities are indicated by the arrows to the right.
      The calculated line luminosities saturate at distances of roughly
      100\,R$_{*}$.
      Beyond this point, the disk parameters, especially $T_{\rm e}$,
      might change, i.e. the disk might cool rapidly, to allow for molecule
      and dust formation at larger distances.

%\begin{figure}[!ht]
%\plotone{smc.eps}
%\caption{Calculated increase in [O{\sc i}] line
%          luminosities with radial distance for the
%          different SMC B[e] supergiants.
%          The arrows indicate the observed values.}
%\end{figure}

\section{The LMC star R\,126}

        R\,126 is different from the SMC B[e] supergiants in the sense
        that it also shows forbidden emission lines of slightly higher
        ionized elements, like [N{\sc ii}], [S{\sc ii}], and [O{\sc ii}].
        In Fig.\,3 we show the results of the fitting of the oxygen lines
        with two different scenarios: (i) the 
        [O{\sc ii}] emission comes from the polar wind only, while the
        [O{\sc i}] emission comes from a completely (hydrogen) neutral disk
        (left panels), and (ii) the [O{\sc ii}] emission comes from the
        still ionized inner parts of the outflowing disk while the [O{\sc i}] 
        emission comes from distances in the disk where O{\sc ii} has already 
        recombined (right panels). Which scenario is really the correct one can 
        only be disentangled with a full analysis of the emission lines
        in combination with the modeling of the continuum emission (ff-fb),
        which is currently in preparation.

\begin{figure}[!ht]
\plotone{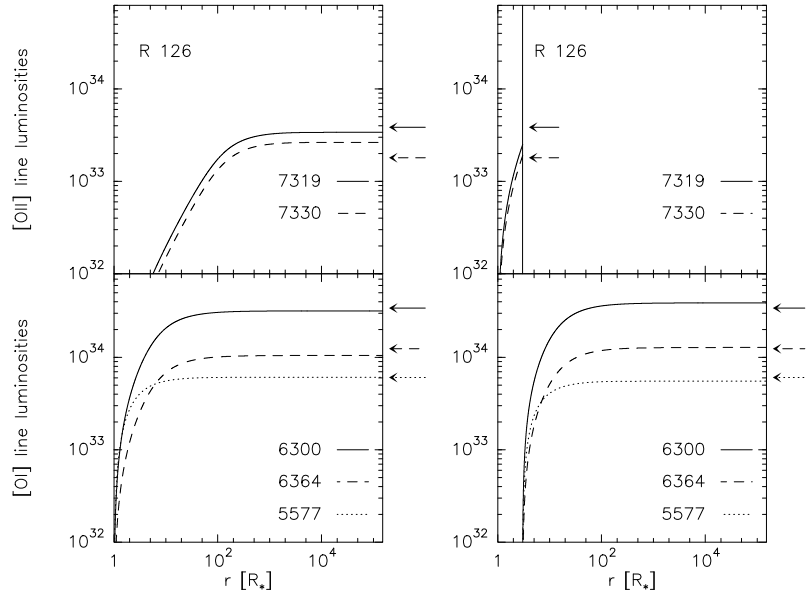}
\caption{Calculated increase in line luminosities
            of the [O{\sc i}] and [O{\sc ii}] lines from the LMC B[e]
            supergiant R\,126 using two different scenarios.
            Left panels: [O{\sc i}] emission from
            the H neutral disk and [O{\sc ii}] emission from the polar wind.
            Right panels: [O{\sc ii}] emission from the inner 3\,R$_{*}$
            of the disk,
            and [O{\sc i}] emission from distances beyond. The observed
            values are indicated with arrows.}
\end{figure}

\vspace{-0.5cm}

\section{Conclusions}

\begin{itemize}
                                                                                
\item  $[$O{\sc i}] emission lines trace the dense and cool
       H neutral disk material around B[e] supergiants.
                                                                                
\item  Modeling of their line luminosities under the assumption of an
       outflowing disk scenario requests that the disk must be neutral
       very close to the stellar surface, in agreement with model
       calculations \citep{KL2003,KL2005}.
                                                                                
\item  The assumptions made for the model calculations are such that
       the disk mass loss rates are {\it lower limits}.
                                                                                
\item  The knowledge of the proper ionization structure of the B[e]
       supergiant stars' disks 
       is very important when modeling the near-IR excess, i.e.
       the free-free and free-bound emission from the non-spherical wind.
                                                                                
\end{itemize}

%\subsection{}   %%% Second level section head (remove "%" symbol)
%\subsubsection{}   %%% Lowest level section head (remove "%" symbol)
%\section*{}	%%% Unnumbered top level section head (remove "%" symbol)
%\subsection*{}   %%% Unnumbered second level section head (remove "%" symbol)

\acknowledgements %%% Text of acknowledgements runs on after this command.
M.K. acknowledges financial support from the Nederlandse Organisatie voor 
Wetenschappelijk Onderzoek grant No.\,614.000.310. M.B.F. acknowledges 
financial support from \emph{CNPq} (Post-doc position - 150170/ 2004-1), 
\emph{Utrecht University}, \emph{LKBF} and \emph{NOVA} foundations.

%%% THE BIBLIOGRAPHY
%%%
%%% CONSULT SECTION 3 OF "INSTRUCTIONS FOR AUTHORS" FOR HOW TO USE NATBIB.
%%% AUTHORS ARE ENCOURAGED TO USE EITHER THE "THEBIBLIOGRAPY" ENVIRONMENT
%%% BY UNCOMMENTING (DELETING THE "%" SYMBOL) THE COMMANDS BELOW, OR BY
%%% USING THE BIBTEX ENVIRONMENT. TO FIND OUT WHICH IS APPLICABLE TO YOUR
%%% CONTRIBUTION, CONSULT THE VOLUME EDITORS FOR YOUR PROCEEDINGS.
%%%

\end{document}